\documentclass[a4paper,11pt]{article}

\usepackage{tikz}
\usepackage{graphicx,array}
\usepackage{color}
\usepackage{latexsym}
\usepackage{feynmp-auto}
\usepackage{amsthm}
\usepackage{amsmath}
\usepackage{amssymb}
\usepackage{empheq}

\usepackage{dsfont}
\usepackage{epsfig}
\usepackage{slashed}
\usepackage{bbold}
\usepackage{psfrag}
\usepackage{subcaption}
\usepackage{xfrac}
\usepackage{multirow}
\usepackage{booktabs}
\usepackage{cite}
\usepackage[normalem]{ulem}
\usepackage{jheppub} % for details on the use of the package, please see the JHEP-author-manual

\newcommand{\be}{\begin{equation}}
\newcommand{\ee}{\end{equation}}
\newcommand{\bea}{\begin{eqnarray}}
\newcommand{\eea}{\end{eqnarray}}
\newcommand{\bei}{\begin{itemize}}
\newcommand{\eei}{\end{itemize}}

\newcommand{\lag}{{\cal L}}
\newcommand{\nn}{\nonumber}

\title{Neutrino dipole portal at a high energy $\mu-$collider}

\author[a,b]{Daniele Barducci and Alessandro Dondarini}
\affiliation[a]{Universit\`a di Pisa and INFN Section of Pisa, Largo Bruno Pontecorvo 3, 56127, Pisa, Italia}
\emailAdd{daniele.barducci@pi.infn.it}
\emailAdd{alessandro.dondarini@phd.unipi.it}

\abstract{We study the phenomenology of $d=6$ dipole portal operators connecting active and sterile neutrinos at a futuristic muon collider. These operators can be the dominant portal between the Standard Model and the New Physics sector in scenarios in which the active-sterile mixing is suppressed. We identify two production modes for sterile neutrinos: one proceeding through the exchange of an $s-$channel electroweak boson and one arising from the fusion of an electroweak boson with a Standard Model lepton. We study the expected reach on the operators suppression scale for these different production mechanisms, showing that the latter offers the best sensitivity and allowing to test a New Physics scale in the $\sim 10\;$TeV range for strongly coupled UV completions of the dipole operators.
}

\begin{document} 
Last update \today
\maketitle

\section{Introduction}\label{sec:intro}

The observational evidences that neutrinos have mass~\cite{ParticleDataGroup:2022pth} imply that the Standard Model (SM) needs to be extended with additional degrees of freedom. Arguably, the simplest possibility is the addition of right-handed (RH) fermions, singlet under the SM gauge group and commonly dubbed {\it sterile neutrinos}. Other than Yukawa interactions with the SM lepton doublets, these states can also have a Majorana mass term and neutrino masses might then be generated by the so-called {\it see-saw mechanism}~\cite{Minkowski:1977sc,Mohapatra:1979ia,Yanagida:1979as,Gell-Mann:1979vob,Schechter:1981bd}. In this case, both the active and sterile neutrinos eigenstates are self-conjugate Majorana fermions. Their masses $m_\nu$ and $m_N$ are schematically related by the see-saw condition
\be
m_\nu \simeq \frac{y_\nu^2 v^2}{m_N} \ ,
\ee
with $y_\nu$ and $v$ being the Yukawa interaction and the Higgs vacuum expectation value. Within this framework sterile neutrinos also inherit couplings to the SM gauge bosons, which are proportional to the active-sterile mixing angle $\theta$ parametrically given by
\be\label{eq:seesaw2}
\theta \simeq \sqrt{\frac{m_\nu}{m_N}} \ .
\ee
For sterile neutrinos with mass at the electroweak (EW) scale the mixing angle is greatly suppressed and direct experimental tests are challenging.\footnote{For $n>1$ sterile neutrinos 
the relation of Eq.~\eqref{eq:seesaw2} turns out to be modified. The mixing angle can receive an exponential enhancement with respect to the naive see-saw scaling case. This is best seen in the Casas-Ibarra parametrization~\cite{Casas:2001sr}. From the practical point of view, this means that masses and mixings can be treated as independent parameters, thus enlarging the experimental accessible parameter space.} While the see-saw extension of the SM can be considered a full-fledged ultraviolet (UV) complete theory, at least in the same way as the SM is, in recent years more and more attention has been devoted to investigate the possibility that it might just be an effective field theory (EFT) description and that higher dimensional operators appearing at a scale $\Lambda$ above, but not too far from, the EW scale might modify its predictions. The resulting theory, usually called $\nu$SMEFT, is described by
\be\label{eq:nusmeft}
\lag  = \lag_{\rm SM} + i \bar N \slashed\partial N - \bar L Y_\nu \tilde H N - \frac{1}{2}\bar N^c M_N N + \sum_{n>4}\frac{{\cal O}^n}{\Lambda^{n-4}} + h.c. \ ,
\ee
where $N$ is a vector describing ${\cal N}_f$ flavors of RH neutrino fields and $N^c = C \bar N^T$ with $C=i \gamma^2 \gamma^0$.
Furthermore $Y_\nu$ is the $3 \times {\cal N}_f$ Yukawa matrix of the neutrino sector with $\tilde H = i \sigma^2 H^*$, $\sigma^2$ the second Pauli matrix while $M_N$ is a  
${\cal N}_f\times {\cal N}_f$
Majorana mass matrix for the RH neutrinos. Finally ${\cal O}^n$
are the Lorentz and gauge invariant operators with mass dimension $n$ built out from the SM and
the RH neutrino fields, with $\Lambda$ parametrizing the scale at which these operators are generated. A
complete and independent set of operators for the $\nu$SMEFT has been built up to $d=9$~\cite{delAguila:2008ir,Liao:2016qyd,Li:2021tsq}.
At the lowest possible dimension $d=5$ only two operators are present: one which couples the RH neutrinos with the Higgs boson $|H|^2\bar N^c_i N_j$ and a dipole interaction with the hypercharge gauge boson
$ \bar N^c_i\sigma^{\mu\nu} N_j B_{\mu\nu}$, which vanishes unless $i\ne j$. These two operators trigger new decay modes for the Higgs and the $Z$ bosons, which can be efficiently tested in $h$ and $Z$ precision measurements as well in high-intensity beam dump experiments, see {\it e.g.}~\cite{Aparici:2009fh,Caputo:2017pit,Barducci:2020icf,Barducci:2022gdv,Delgado:2022fea,Duarte:2023tdw,
Chun:2024mus}

On the other-side at $d=6$ many other operators appear. Among these there are two dipole interactions
$\bar L \sigma^{\mu\nu} N \tilde H B_{\mu\nu}$ and $\bar L \sigma^{\mu\nu} N \sigma^a \tilde H W_{\mu\nu}^a $ which at low energy appear as a photon dipole portal between active and sterile neutrinos $\bar \nu \sigma^{\mu\nu} N F_{\mu\nu}$. These operators, that can be the dominant portal between the SM and the new physics (NP) sector in the case that the active-sterile neutrino mixing is highly suppressed, can be efficiently tested in various ways, see {\it e.g.}~\cite{Magill:2018jla,Brdar:2020quo,Ismail:2021dyp,Zhang:2022spf,Huang:2022pce,Barducci:2023hzo,Zhang:2023nxy,Ovchynnikov:2023wgg}. For sub-GeV sterile neutrino masses, high-intensity fixed target experiments can exploit the large production rates of QCD mesons and search for their exotic decays, while astrophysical probes can further extend the reach to lower masses. For higher masses above the GeV, collider searches at LEP, LHC and future $e^+e^-$ facilities offer the most important probes. 

Sterile neutrinos with masses around the TeV scale are 
a challenging target for the LHC~\cite{CERN-EP-2024-083}
 and could be an ideal target for a future proposed $\mu-$collider ($\mu$C) operating with at  multi-TeV center of mass (COM) energies~\cite{AlAli:2021let,Aime:2022flm,Accettura:2023ked}. The phenomenology of sterile neutrinos SM extensions has begun to be explored at $\mu$C for what concerns its renormalizable form~\cite{Li:2023tbx,Mekala:2023diu,Kwok:2023dck,Li:2023lkl} as well as for $d=6$ four-fermi operators~\cite{Barducci:2022hll}, with the motivation for the latter case that one could exploit the growth with the energy of the scattering amplitudes.~\footnote{For related works on the LHC phenomenology of four-fermi operators see, {\it e.g.},~\cite{Duarte:2019rzs,Cottin:2021lzz,Beltran:2022ast,Mitra:2022nri,Beltran:2023ksw}}

The large COM energy of a $\mu$C has however also another benefit, in that it allows to effectively scatter EW gauge bosons radiated from the colliding $\mu$ beams~\cite{Costantini:2020stv,Han:2020uid,Ruiz:2021tdt,Garosi:2023bvq}. The goal of this study is then to explore if this feature can be exploited to efficiently test sterile neutrinos dipole operators that connect the active and sterile sector via an EW gauge boson. In order to pin down the relevant feature of the dipole operators signatures, we will work under the assumption that the active-sterile mixing plays a subdominant role to the sterile neutrino phenomenology.

The rest of the paper is organised as follows. In Sec.~\ref{sec:th} we define our theoretical framework and in Sec.~\ref{sec:muC} we describe the main characteristics of the future $\mu$C 
prototypes that we consider. We start then in Sec.~\ref{sec:warm} with a warm-up analysis by only considering the dipole interaction in the EW broken phase, in order to illustrate the most important signatures and signal characteristics that we will subsequently study in the full fledged SM gauge invariant formalism in Sec.~\ref{sec:SM_gauge}. We then summarise our findings and conclude in Sec.~\ref{sec:concl}.

\section{Theoretical framework}
\label{sec:th}

We extend the SM with a single RH Majorana fermion $N$ and work under the assumption of negligible mixing between the sterile and the active neutrinos.
Below the EW scale the leading portal between the SM and the NP sector is parametrized by the $d=5$ dipole operator
\be\label{eq:IR_dipole}
\lag = d_{\gamma}^i \bar \nu_L^i \sigma^{\mu\nu} N F_{\mu\nu} + h.c. \ , \qquad i = e, \mu, \tau \ ,
\ee
where $2\sigma^{\mu\nu}=i[\gamma^\mu,\gamma^\nu]$
and $F_{\mu\nu}$ is the photon field strength tensor. The operator of Eq.~\eqref{eq:IR_dipole} doesn't however respect the SM gauge symmetries and above the EW scale must be described by full SM gauge invariant interactions which can be parametrized as
\be\label{eq:dipole_d6}
\lag = \bar L^i \left( {\cal C}_{\cal W}^i W_{\mu\nu}^a \frac{\sigma^a}{2} + {\cal C}_{\cal B}^i B_{\mu\nu}\right)\tilde H \sigma^{\mu\nu} N + h.c. \ ,
\ee
where $\sigma^a$ are Pauli matrices and $B_{\mu\nu}$ and $W_{\mu\nu}^a$ the field strength tensors of the $U(1)_Y$ and $SU(2)_L$ gauge bosons respectively.
After EW symmetry breaking (EWSB) one obtains the following $d=5$ operators
\be\label{eq:dipole_UV_IR}
\lag = 
d_Z^i \bar\nu_L^i \sigma^{\mu\nu} N Z_{\mu\nu} + 
d_\gamma^i \bar\nu_L^i \sigma^{\mu\nu} N F_{\mu\nu} + 
d_W^i \bar e_L^i \sigma^{\mu\nu} N W^-_{\mu\nu} +  h.c. \ ,
\ee
with
\begin{align}
\label{eq:dipole_UV}
d_\gamma^i & = \frac{v}{\sqrt 2}\left(\frac{{\cal C}_{\cal W}^i}{2} s_\omega + {\cal C}_{\cal B}^i c_\omega \right) \ , \nn \\
d_Z^i & = \frac{v}{\sqrt 2}\left(\frac{{\cal C}_{\cal W}^i}{2} c_{\omega} - {\cal C}_{\cal B}^i s_\omega \right) \ , \nn \\
d_W^i & = \frac{v}{2} {\cal C}_{\cal W}^i  \ , 
\end{align}
with $\omega$ indicating the Weinberg angle,
so that only one of these three operators at the time can be switched off by a proper choice of the $d=6$ Wilson coefficients.\footnote{
Note that 
the operator of Eq.~\eqref{eq:IR_dipole} can also arise in theories with $n>1$ RH neutrinos from 
a $d=5$ dipole as $ d_{\cal N} \bar N_{1,R}^c \sigma^{\mu\nu}  N_{2,R} B_{\mu\nu} + h.c.
$
in the presence of a mixing between the active and sterile sector.
Since we work under the hypothesis of negligible active-sterile mixing, we don't consider this possibility
further.
} In the following we will assume for concreteness that the Wilson coefficients ${\cal C}_{\cal W}^i$ and ${\cal C}_{\cal B}^i$ are real.
As a warm-up study, and in order to highlight the most important phenomenological features and signal characteristics that can appear at a $\mu$C, in the following section we will analyse the phenomenology of the single $d_\gamma^i$ operator working at $d=5$, while we defer to Sec.~\ref{sec:SM_gauge} the analysis of the full $d=6$ SM gauge invariant formulation of Eq.~\eqref{eq:dipole_d6}.

\section{$\mu-$collider specifications}
\label{sec:muC}

In recent years a collider that employs high-energy muon beams has become a concrete possibility for the post LHC era. If realized, it will enable the direct study of physics in multi-TeV
regime. It has been estimated than a $\mu$C with a COM energy of $\sim 10\;$TeV can achieve the same rates for the production of EW beyond the SM (BSM) states comparable to the one that can be attained at a 100\;TeV $pp$ collider~\cite{Aime:2022flm}, which is another proposed project under close scrutiny at the moment~\cite{FCC:2018byv,FCC:2018vvp,Bernardi:2022hny}. The $\mu$C option has however also the advantage of a cleaner experimental environment, thus leveraging on the unique possibility of combining a direct exploration of the multi-TeV regime with very precise measurements. Moreover, it might be realized within a timescale shorter than the 100\;TeV $pp$ option.

The actual feasibility of a $\mu$C is however yet to be established as many technological challenges must be overcome in order to firmly establish whether this dream machine can be realized. Nevertheless  the high-energy community is already actively analysing what could be its potentiality in terms of SM and BSM studies, see {\it e.g.}~\cite{AlAli:2021let,MuonCollider:2022xlm,Accettura:2023ked} for reviews.
The proposed $\mu$C COM energies range from $\sqrt s=m_h$, to allow for extremely precise Higgs physics analyses, up to multi-TeV prototypes with $\sqrt s$ as large as ${\cal O}(30\,$TeV)~\cite{AlAli:2021let}.
In order to compensate for the $1/s$ scaling of the cross-section of $s-$channel processes, the target integrated luminosity that has been so far considered is assumed to scale as
\be\label{eq:lumi}
{\cal L} = 10\;{\rm ab}^{-1}\left(\frac{\sqrt s}{10\;{\rm TeV}}\right)^2 \ ,
\ee
which is claimed to be possibile given the characteristics of the $\mu$ beams~\cite{Accettura:2023ked}.
In our study we will consider two benchmarks COM energies and luminosities\footnote{For  the case $\sqrt s=3\,$TeV the integrated luminosity that we take is slightly larger than the one given by the relation of Eq.~\eqref{eq:lumi}. This is however a common benchmark in $\mu$C analyses, see {\it e.g}~\cite{MuonCollider:2022xlm}.}:
$\sqrt s=3\,$TeV and ${\cal L}=1\,$ab$^{-1}$ and $\sqrt s=10\,$TeV and ${\cal L}=10\,$ab$^{-1}$. 
Given the preliminary nature of $\mu$C studies, and in order to investigate what can be achieved with smaller or larger integrated datasets, we will also present our results by  considering the integrated luminosity as a free parameter for the two chosen COM energies.

An important feature of a multi-TeV $\mu$C is 
that at high-energy hard processes initiated by vector boson fusion (VBF)  can dominate over the ones initiated by direct $\mu^+\mu^-$ annihilation for both SM and BSM processes~\cite{Costantini:2020stv,Han:2020uid,Ruiz:2021tdt,Garosi:2023bvq}, given the high probability of radiating an EW gauge boson from the incoming $\mu-$beam.
For what concerns our investigations, this means that sterile neutrinos with dipole interactions might efficiently be produced in hard-scattering processes involving a SM EW gauge boson and a lepton. This turns out to be the production mechanism offering the best sensitivity on this scenario, as we will show throughout our analysis. In order to better illustrate the various sterile neutrino production topologies we will start our study with a warm-up analysis, where we only consider the $d=5$ dipole portal with the photon described by Eq.~\eqref{eq:IR_dipole}, moving then in Sec.~\ref{sec:SM_gauge} to the more appropriate $d=6$ SM gauge invariant formulation.

\section{Warm-up: photon dipole with exclusive mixing with the $\mu$ flavor}
\label{sec:warm}

The $d=5$ dipole operator of Eq.~\eqref{eq:IR_dipole} triggers the decay of the sterile neutrino into an active neutrino and a photon with a rate
\be\label{eq:d5_decay}
\Gamma_{N\to \nu^i \gamma} = \frac{{d_\gamma^i}^2}{2\pi}m_N^3 
\ ,
\ee
with $i=e,\mu\,\tau$ parametrizing the direction in flavor space.
Depending on the values of $d^i_\gamma$ and $m_N$, the sterile neutrino might decay promptly, in proximity of the $\mu$ beams interaction point, or else have a longer lifetime and give rise to displaced decays or fully invisible signatures. However, as we shall see, for the range of parameters of interest for this study, the sterile neutrino never exhibits a displaced decay pattern, and we will then restrict our analysis to the study of standard prompt like signatures.

\begin{figure}[t!]
\begin{center}
\includegraphics[width=0.38\textwidth]{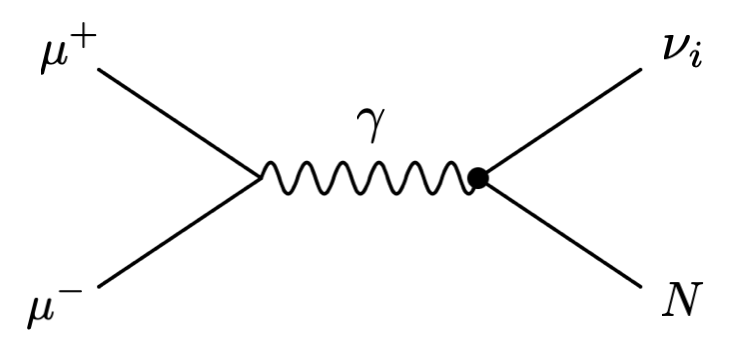}
\hspace{2cm}
\includegraphics[width=0.33\textwidth]{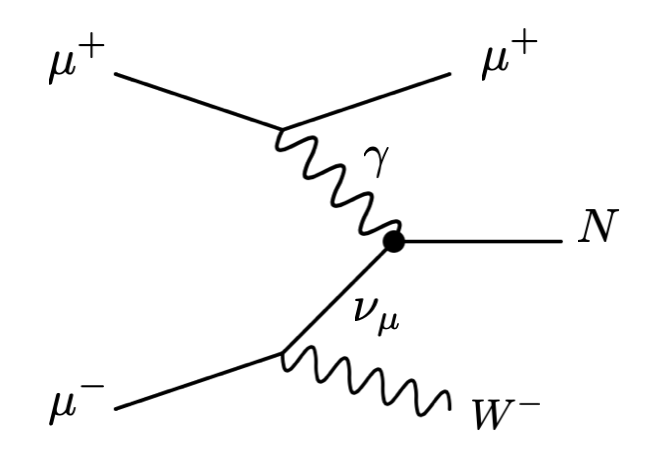}
\caption{
Representative Feynman diagrams for $s-$channel (left) and $\gamma \nu$ fusion (right) single $N$ production via the dipole operator of Eq.~\eqref{eq:IR_dipole}. The black dots correspond to an insertion of the effective operator.
}
\label{fig:single_prod_IR}
\end{center}
\end{figure}

\begin{figure}[t!]
\begin{center}
\includegraphics[width=0.32\textwidth]{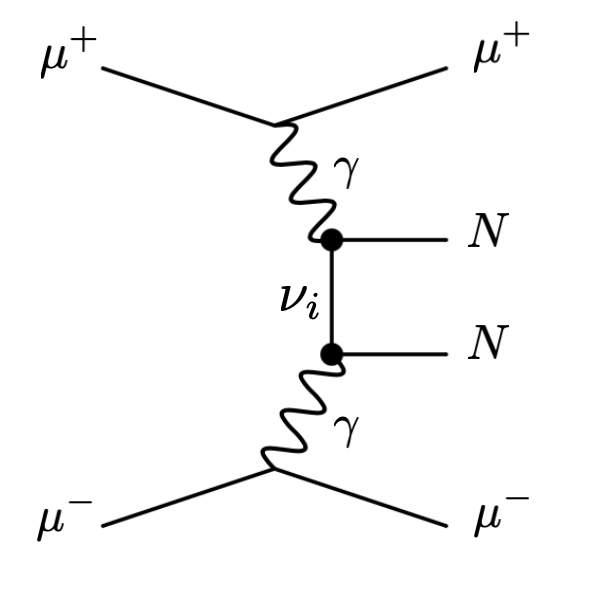}
\hfill
\caption{Representative Feynman diagram for VBF pair $N$ production via the dipole operator of Eq.~\eqref{eq:IR_dipole}. The black dots correspond to an insertion of the effective operator.}
\label{fig:double_prod_IR}
\end{center}
\end{figure}

As concerning its production modes, there are different ways in which the sterile neutrino can be produced at a $\mu$C. From one side they can be singly or pair-produced. Single production occurs through the following processes: production through the $s-$channel exchange of a photon and $\gamma \nu$ fusion, whose topologies are depicted in Fig.~\ref{fig:single_prod_IR}.
The $\gamma \nu$ fusion process exploits the high-probability for the colliding $\mu$ beam to radiate an EW gauge boson.\footnote{These processes can be efficiently described with the language of $\mu-$parton distribution functions (PDFs)~\cite{Costantini:2020stv,Han:2020uid,Ruiz:2021tdt,Garosi:2023bvq}. For our analysis we will however rely on the generation of the full matrix element for the processes under investigation.}
Note that while the $s-$channel production is present for all flavor assignments of the $d_\gamma^i$ coefficients, the $\gamma \nu$ fusion only occurs when the sterile neutrino interacts with the $\mu$ flavor.\footnote{
Interactions with the $e$ and $\tau$ flavor can be tested by higher order diagrams or, in the language of $\mu$ parton distributions, by considering the $\nu_L^{e,\tau}$ PDFs of the muon. We find these contributions to be negligible.
} To highlight the various features of these two production modes, in this section we will only switch on the $d_\gamma^\mu$ Wilson coefficient.  On the other side, sterile neutrinos can also be pair-produced through the VBF diagram of Fig.~\ref{fig:double_prod_IR}, which allows to test all the $d_\gamma^i$ flavor assignments. This diagram presents however a double operator insertion that suppresses its production rate. 

In order to analyse the neutrino dipole portal model, we have implemented the  operators of Eq.~\eqref{eq:IR_dipole} and Eq.~\eqref{eq:dipole_d6} in the {\tt UFO}~\cite{Degrande:2011ua} format through the {\tt Feynrules}~\cite{Alloul:2013bka} package 
and used {\tt MadGraph5\_aMC\@NLO}~\cite{Alwall:2014hca} as event generator. 
%%%
Details for the implementation of the Majorana fermions interactions are reported in
App.~\ref{app:MG}.
%%%
The single  $N$ production processes have a single dipole insertion. After the sterile neutrino has decayed into a $\gamma \nu$ final state~\footnote{Note that before the decay of $N$ the process under consideration is $\mu W N$ production, which can also occur via active-sterile mixing, see {{\it e.g}}~\cite{Antonov:2023otp}.}, for the $s-$channel production mechanism 
the experimental signature is the one of a 
highly-energetic photon and missing energy. In the $\gamma \nu$ fusion process the same final state is also accompanied by the remnant of the incoming $\mu$ beams splitting, which are a muon and a $W$ boson. As we will show later the former is generically forward, {\it i.e.} with a small polar angle with respect to the beam axis, while the latter is more central. 
Both these features can in principle be used to tag this specific topology and possibly enhance the sensitivity on this scenario.
 The $N$ pair-production proceeds instead via the $\gamma\gamma$ fusion process of Fig.~\ref{fig:double_prod_IR}. The final state presents two photons, missing energy and two muons, which are expected to be predominantly  in the forward direction.
 
\begin{figure}[t!]
\begin{center}
\includegraphics[width=0.45\textwidth]{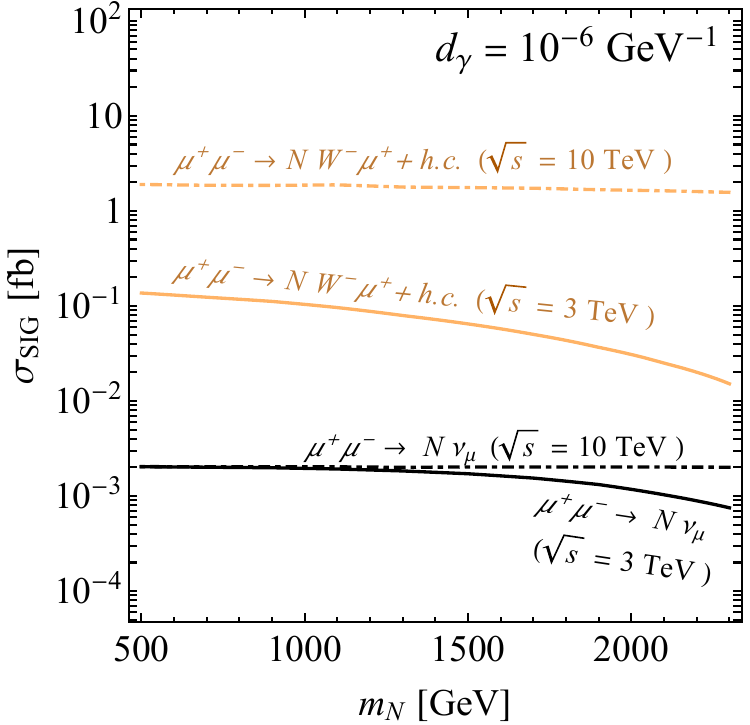}
\hfill
\includegraphics[width=0.47\textwidth]{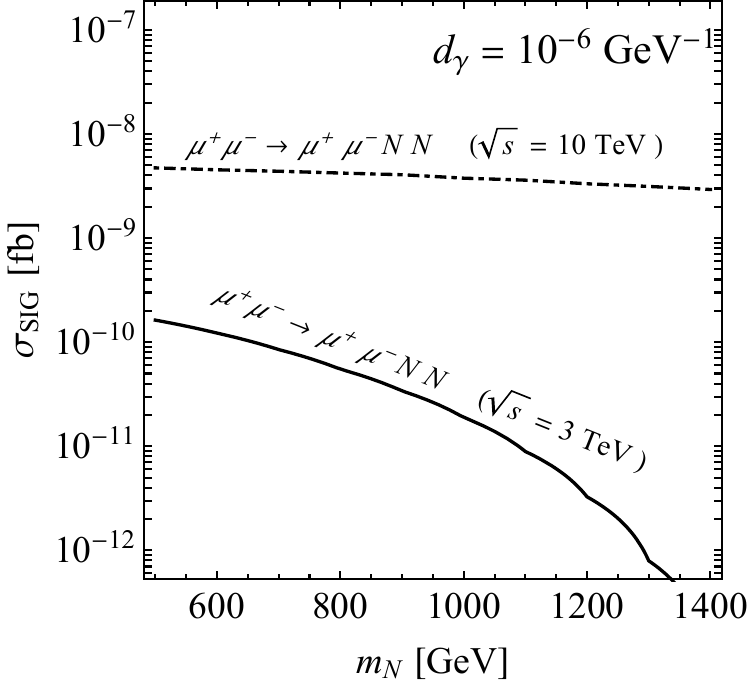}
\caption{
{\it Left:} inclusive cross-section for sterile neutrino single production through the diagrams of Fig.~\ref{fig:single_prod_IR}. We show separately the contributions arising from the $s-$channel and $\gamma \nu$ fusion topologies.
{\it Right:} inclusive cross-sections for sterile neutrino pair production through the VBF diagram of Fig.~\ref{fig:double_prod_IR}. 
In both plots we choose $\sqrt s = 3\;$TeV and $\sqrt s = 10\;$TeV and
the rates are normalized to $d_{\gamma}=10^{-6}\, \text{GeV}^{-1}$.
}
\label{fig:sigxsec}
\end{center}
\end{figure}

 In Fig.~\ref{fig:sigxsec} we show the inclusive cross-sections for these production modes  in function of the sterile neutrino mass for the chosen COM benchmarks, namely $\sqrt s=3\;$TeV and 10\;TeV, by fixing $d_\gamma^\mu=10^{-6}\;$GeV$^{-1}$. The rates for different values of the $d_\gamma^\mu$ coefficient can be inferred from the ${d_\gamma^\mu}^2$ and ${d_\gamma^\mu}^4$ scalings for single and pair-production respectively.
It's interesting to note that the $s-$channel process has a cross-section that is almost two orders of magnitude below the $\gamma \nu$ fusion one. We might then expect a higher sensitivity on the NP parameter space by exploiting this production mode with respect to the $s-$channel production.

For our analysis we work at parton level and make the following approximations for what concerns the efficiencies for the identification and reconstruction of final state objects, which we also apply to the analysis of Sec.~\ref{sec:SM_gauge}.
 As regarding final state photons and muons
 we consider the study~\cite{MuonCollider:2022ded} performed by the
{\it{International Muon Collider Collaboration}} at $\sqrt s=1.5~$TeV and assume that their results can be extrapolated to higher $\sqrt s$.
We further neglect the effects of beam-induced backgrounds, which will slightly degrade the resulting limits that can thus be interpreted as an optimal reach for the proposed analyses.
We fix the photon reconstructed efficiency to be 100\%, 
which is a good approximation for the range of energies and polar angles of interest of this work. Still from~\cite{MuonCollider:2022ded}, the reconstruction efficiency for final state muons in the central part of the detector is found to be around 95\%. 
For what concerns muons in the forward part of the detector which 
we will employ in our analysis to tag the fusion topology as we will discuss later, to the best of our knowledge no experimental dedicated studies are yet available. There exist however a preliminary version of a {\tt Delphes} detector card\,\footnote{The {\tt{Delphes}} card can be found at the following web-page \url{https://github.com/delphes/delphes/blob/master/cards/delphes_card_MuonColliderDet.tcl}} which sets as baseline reconstruction efficiency for forward muons at the detector level a value of 95\%. 
We thus decide to apply a flat $\varepsilon^{\rm reco}_{\mu}=0.95$ reconstruction efficiency for muons in all the kinematic range of interest. 
Still from~\cite{MuonCollider:2022ded} we fix to be a flat 95\% also the efficiency reconstruction for electrons.
We do not consider charge mis-identification rates.
EW gauge bosons are instead treated as follows. We are interested in the high-energy regime, where these object  
 are highly boosted, and we treat them as fat jets. We do not simulate their decays, but instead consider their dominant hadronic decay channels. Concretely, we multiply the signal and background event yields by the relevant branching ratios (BRs), BR($W^\pm \to jj)\simeq 67\%$ and BR($Z \to jj)\simeq 70\%$. No estimates for the reconstruction efficiencies of EW gauge bosons at a $\mu$C is however yet available and we then decide to conservatively use public LHC results derived by the CMS collaboration and reported in
~\cite{CMS-DP-2023-065} as also recently done in similar works, see {\it e.g.}~\cite{Liu:2023jta}.  In particular we apply the identification and reconstruction probabilities 
$\varepsilon_W^{\rm reco} = 0.73$ and $\varepsilon_Z^{\rm reco} = 0.64$.
We also consider the following mis-tag probabilities 
$\varepsilon_{W\to Z}^{\rm mis-tag}=0.17$ and $\varepsilon_{Z\to W}^{\rm mis-tag}=0.22$ setting to zero all the others.
Although full simulations of the detector performances, when they will be available, will certainly allow for more accurate and robust projected sensitivities, we are confident that in a more realistic approach the excellent accuracy of particle-flow reconstruction, under consideration for a $\mu$C~\cite{Accettura:2023ked}, complemented by advanced analyses techniques, might only moderately degrade the present sensitivity estimates.

\begin{figure}[t!]
\begin{center}
\includegraphics[width=0.45\textwidth]{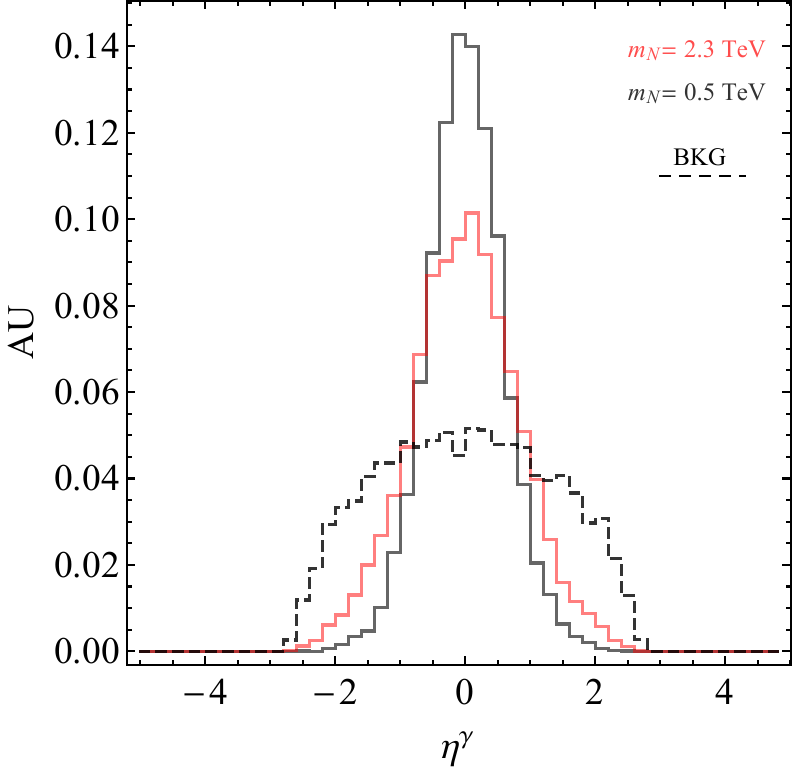}
\hfill
\includegraphics[width=0.45\textwidth]{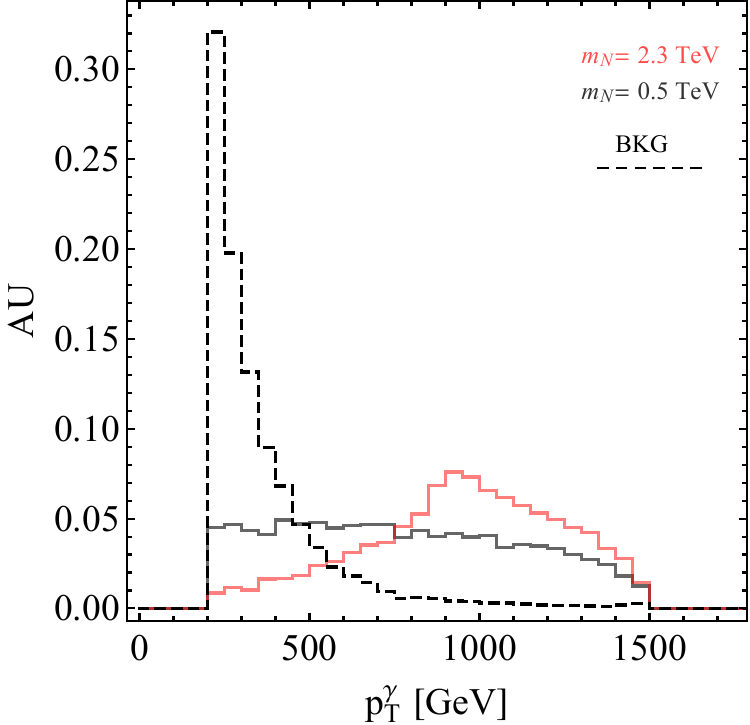}
\caption{
Normalized $\eta^{\gamma}$ (left) and $p_T^\gamma$ (right) events distributions for the $s-$channel signal and the corresponding SM background for $\sqrt s =3\;$TeV and two representative choices of $m_N$.}
\label{fig:dist_sch}
\end{center}
\end{figure}

The dominant background for the $s-$channel  production, after $N$ has decayed,  is given by the SM process $\mu^+\mu^- \to \gamma \nu \bar \nu$\footnote{Additional processes such as $\mu^+\mu^- \to \gamma \; n \times(\nu \bar \nu)$ with $n>1$ are coupling and phase space suppressed while processes with multiple non reconstructed leptons as $\mu^+ \mu^- \to \gamma \ell^+ \ell^-$ are found to give a negligible contribution.}, while for the $\gamma \nu$ fusion process is $\mu^+ \mu^- \to \gamma \mu^+ W^- \nu_\mu + h.c.$. Additional backgrounds coming from multiple mis-identified and/or non reconstructed objets, such as $\gamma \mu^+ Z\mu^-$ where one lepton is lost and the $Z$ is mis-identified for a $W$, are found to be negligible. These considerations will also apply for the analyses of Sec.\,\ref{sec:SM_gauge}, where we will then only consider true backgrounds, with the same final states of the selected signals, rescaled by the relevant reconstruction efficiencies and hadronic BRs for the EW bosons.
 We simulate both signal and background events at leading order with ${\tt MadGraph}$. For the $s-$channel processes we apply the generator level cut $|\eta^\gamma|<3$ and $p_T^\gamma>200\;$GeV ($p_T^\gamma>1000\;$GeV) for the $\sqrt{s}=3\;$TeV ($\sqrt{s}=10\;$TeV ) cases, where these choices have been made in order to optimise the Monte Carlo statistics, since the $\gamma$ from the $N$ decay is expected to be central and highly energetic, given the range of $N$ masses in which we are interested in. For these generator level cuts the background cross-sections are $\sim 204\;$fb for $\sqrt s =3\;$TeV and  $\sim 18\;$fb for $\sqrt s=10\;$TeV.  We then show in Fig.~\ref{fig:dist_sch} the normalized events distributions
for the photon pseudorapidity $|\eta^\gamma|$ and its transverse momentum $p_T^\gamma$ for both the signal and background processes for the case $\sqrt s =3\;$TeV. From the figures it is clear that for the ranges of $N$ masses of our interest a good background rejection can be obtained by simply tightening the cut on $|\eta^\gamma|$ and $p_T^\gamma$.

\begin{figure}[h!]
\begin{center}
\includegraphics[width=0.45\textwidth]{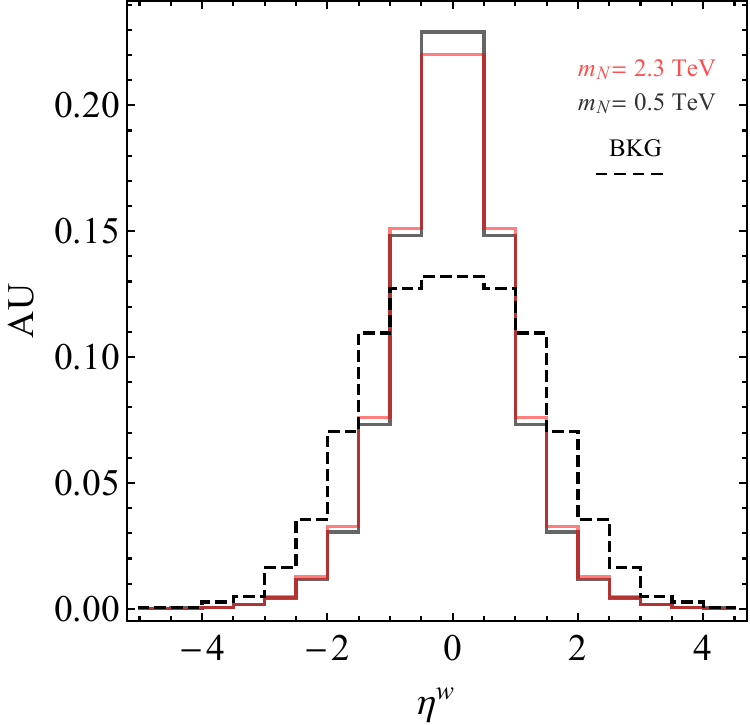}
\hfill
\includegraphics[width=0.45\textwidth]{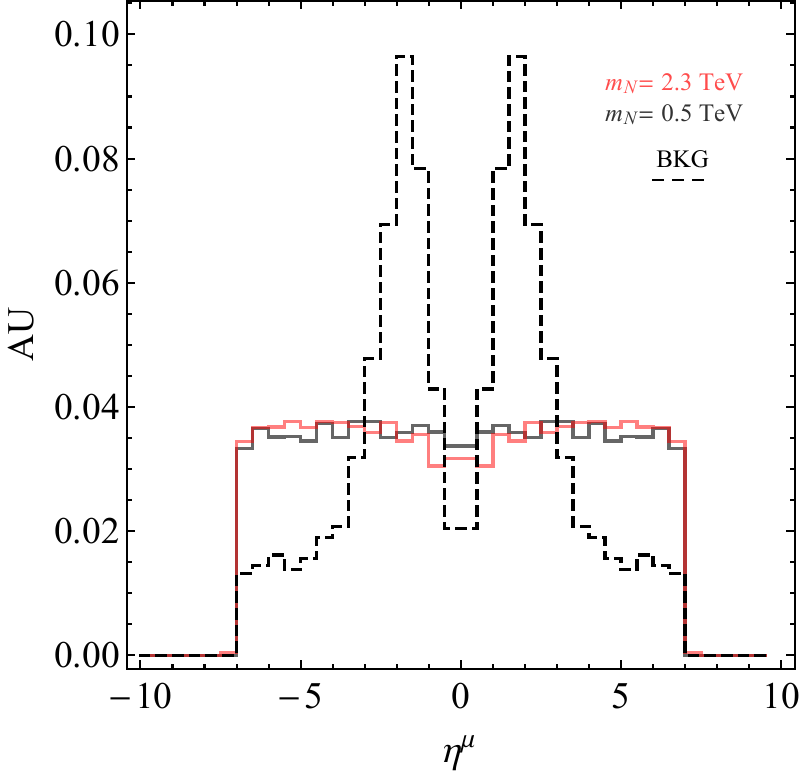}
\caption{
Normalized $\eta^{W}$ (left) and $\eta^\mu$ (right) events distributions for the $\gamma\nu$ fusion signal and the corresponding SM background for $\sqrt s =3\;$TeV and two representative choices of $m_N$.}
\label{fig:dist_VBF}
\end{center}
\end{figure}

Moving now to the $\gamma \nu$ fusion production, here we have imposed at generator level 
the same $|\eta^\gamma|$ and $p_T^\gamma$ requirements as for the $s-$channel case, since the photon arising from the sterile neutrino decay is expected to have similar distributions, {\it i.e.} to be central and highly energetic. This is confirmed by analysing the corresponding distributions, which are similar to the $s-$channel production case and that we therefore do not show. However in the $\gamma\nu$ fusion production process one can have an extra handle in order to discriminate the signal over the background due to the particular signal topology, see the right panel of Fig.~\ref{fig:single_prod_IR}. In particular one expects the final state muon to be peaked at large values of $|\eta^\mu|$, {\it i.e.} emitted at a small angle with respect to the beam line. A dedicated forward muon spectrometer can in principle be installed in the $\mu$C environment, which can be used to tag events where the $\mu$ beams radiate a neutral EW gauge boson that participates in the hard scattering.
This has been firstly proposed in~\cite{Ruhdorfer:2023uea}, where the Authors investigated the prospect of searching for BSM invisible decays of a SM Higgs boson produced through the $ZZ$ VBF fusion process. 

Following the discussion in~\cite{Ruhdorfer:2023uea} we have therefore further applied at generator level the cuts $E_{\mu}> 500$ GeV and $|\eta^\mu|< 7$, where the former selection is imposed in order for the muons to be identified, since with this energy they should be able to 
penetrate the foreseen conical absorbers along the beam line, while the latter is applied in order to restrict the kinematic coverage of the forward muon spectrometer to realistic angles, see again~\cite{Ruhdorfer:2023uea}.
The background cross-section normalization with these generator level cuts are $\sim 10\;$fb for $\sqrt s =3\;$TeV and $\sim 2\;$fb for $\sqrt s =10\;$TeV.
We then show in Fig.~\ref{fig:dist_VBF} the $\mu$ and $W$  pseudorapidity distributions. The two plots show that, while the emitted muon distribution has for the signal a large forward component, as expected in VBF processes, the same is not true for the $W$ radiated from the initial state muon, whose distribution peaks at small  values of $|\eta^W|$.

\begin{table}[h!]
\begin{center}
\scalebox{0.87}{
\begin{tabular}{ c |  c c || c | c c }
\multicolumn{6}{c}{
$\mu$C at $\sqrt s=3\;$TeV~~${\cal L}=1\;$ab$^{-1}$}\\
\hline
 \multicolumn{3}{c||}{{\bf $s-$channel}} & \multicolumn{3}{c}{{\bf 
$\gamma\nu$ fusion}} \\
\hline
Cut 
& $\sigma_{\rm bkg}\;$[fb]
& $\sigma_{\rm sgn}\;$[fb]
& Cut
& $\sigma_{\rm bkg}\;$[fb]
& $\sigma_{\rm sgn}\;$[fb]  \\ 
\hline
Gen. level 
&$ 2.0\times 10^2$
& $1.7 \times 10^{-3}$
& Gen. level
&  10
&  $2.2 \times 10^{-2}$\\
$p_T^\gamma>900\;$GeV
&   5.9
& $6.2 \times 10^{-4}$
& $p_T^\gamma>600\;$GeV,\;$2.44<|\eta^\mu|<7$
& $1.4 \times 10^{-1}$ 
&  $3.4 \times 10^{-3}$
\end{tabular}}
\vskip 20pt
\scalebox{0.87}{\begin{tabular}{ c |  c c || c | c c }
\multicolumn{6}{c}{
$\mu$C at $\sqrt s=10\;$TeV~~${\cal L}=10\;$ab$^{-1}$}\\
\hline
 \multicolumn{3}{c||}{{\bf $s-$channel}} & \multicolumn{3}{c}{{\bf 
$\gamma\nu$ fusion}} \\
\hline
Cut 
& $\sigma_{\rm bkg}\;$[fb]
& $\sigma_{\rm sgn}\;$[fb]
& Cut
& $\sigma_{\rm bkg}\;$[fb]
& $\sigma_{\rm sgn}\;$[fb]  \\ 
\hline
Gen. level 
& 18
& $1.5 \times 10^{-3}$
& Gen. level
&  2.0
&  $2.4 \times 10^{-1}$\\
$p_T^\gamma>3500\;$GeV
&   $3.3 \times 10^{-1}$
& $4.1 \times 10^{-4}$
& $p_T^\gamma>2500\;$GeV,\;$2.44<|\eta^\mu|<7$
&  $2.1 \times 10^{-2}$ 
&  $3.6 \times 10^{-2}$
\end{tabular}}
\end{center}
\caption{
Cut-flow for the $s-$channel and $\gamma\nu$ fusion processes and corresponding backgrounds for the new physics operator of Eq.~\eqref{eq:IR_dipole} at a $\mu$C with $\sqrt s=3\;$TeV (upper table) and $\sqrt s=10\;$TeV (lower table). For the signal, we fix $m_N=500\;$GeV and $d_\gamma=10^{-6}\;$GeV$^{-1}$. See main text for the definition of the generator level cuts.
}
\label{tab:cut_flow_warm_up}
\end{table}

In order to optimize our choices of analysis cuts, we define the statistical significance $z$ as
\be
z = \frac{N_S}{\sqrt{N_S+N_B + \alpha_{\rm sys}^2 N_B^2}} \ ,
\ee
where $N_{S,B}$ are the physical number of signal and background events respectively, while $\alpha_{\rm sys}$ is a systematic uncertainty on the background estimation, which we set to 2\% as a reference value.
We use the value of $z=2$ as an estimate of the 95\% confidence level exclusion. 
\begin{figure*}[t!]
\begin{center}
	\includegraphics[width=0.482\textwidth]{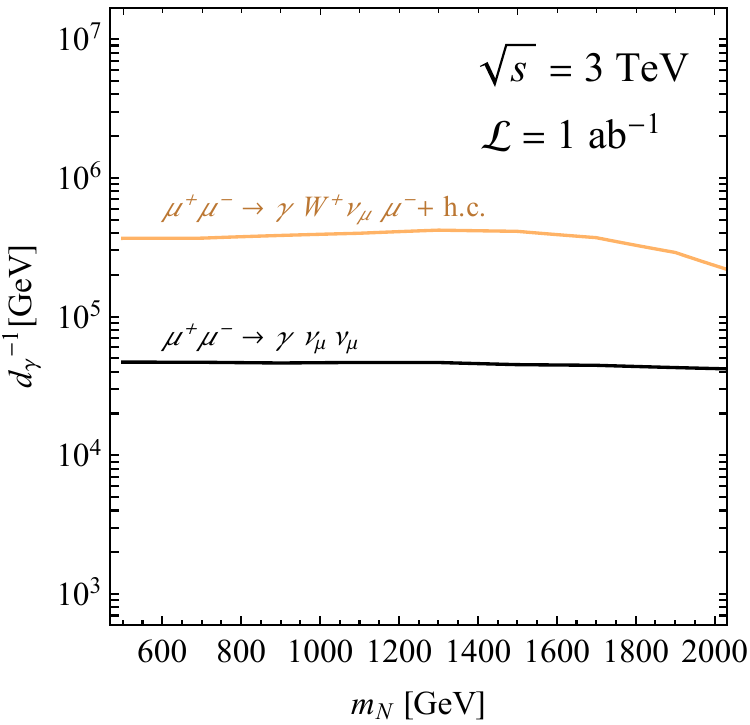}\hfill
 \includegraphics[width=0.482\textwidth]{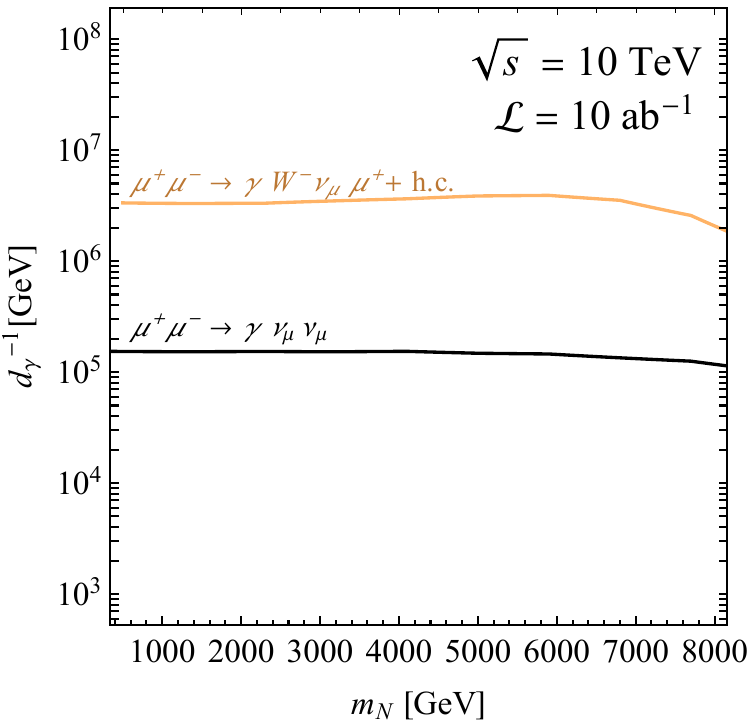}\hfill
 \caption{Expected sensitivity on the new physics operator of Eq.~\eqref{eq:IR_dipole} as a function of the sterile neutrino mass $m_N$
for a $\mu$C with $\sqrt s=3\;$TeV and ${\cal L} = 1\;$ab$^{-1}$ (left) and $\sqrt s=10\;$TeV and ${\cal L} = 10\;$ab$^{-1}$ (right). We show in black and in orange the reach from the $s-$channel and $\gamma\nu$ fusion
processes respectively.}
 \label{fig:reach}
 \end{center}
 \end{figure*}
We then optimise the reach in the $m_N - d_\gamma^\mu$ parameter space by restricting the photon pseudorapidity to $|\eta^\gamma|<2.44$ and its transverse momentum to
$p_T^\gamma>900\;$GeV ($p_T^\gamma>3500\;$GeV) and $p_T^\gamma>600\;$GeV ($p_T^\gamma>2500\;$GeV) for the $s-$channel and $\gamma \nu$ fusion processes for $\sqrt{s}= 3\;$TeV ($\sqrt{s}= 10\;$TeV). In addition, we also have restricted the muon pseudorapidity to $2.44<|\eta^\mu|<7$. We have chosen to optimise the reach for the lowest simulated value of the sterile neutrino mass~\footnote{The acceptance of the transverse momentum cut is higher for higher sterile neutrino masses, since the $p^{\gamma}_T$ distribution shifts to larger energies, while $\eta^{\gamma}$ is pretty insensitive to the $N$ mass.}, namely $m_N=500\;$GeV and we show in Tab.~\ref{tab:cut_flow_warm_up} representative cut-flows for
the signal and background cross-sections fixing $m_N=500\;$GeV and $d_\gamma=10^{-6}\;$GeV$^{-1}$.
The reaches in the $m_N - d_\gamma^\mu$ plane
are then shown in Fig.~\ref{fig:reach} for $\sqrt s=3\;$TeV and an integrated luminosity of ${\cal L} = 1\;$ab$^{-1}$ and $\sqrt s=10\;$TeV and ${\cal L} = 10\;$ab$^{-1}$ in the left and right panel respectively.
In both plots we separately show the limits that can be attained in the $s-$channel and $\gamma\nu$ fusion  processes respectively.
As one can see, by exploiting the $\gamma\nu$ channel allows to greatly improve the sensitivity on $d_\gamma^\mu$ with respect to the one obtained from the $s-$channel. Note also that for the range of masses and NP scales that can be tested, the sterile neutrinos always decays promptly in the detector as anticipated at the beginning of this section, see Eq.~\eqref{eq:d5_decay}. For what concerns the pair-production topology, we do not expect it to play a relevant role in testing this scenario, given its double operator suppression and the consequent smaller cross-section with respect to the $s-$channel and $\gamma\nu$ fusion cases, see Fig.~\ref{fig:sigxsec}.

\section{SM gauge invariant case in flavor space}
\label{sec:SM_gauge}

The previous section has served to illustrate the production modes of possible relevance at a multi-TeV $\mu$C in a simplified framework. However for energies above the EW scale the dipole interactions need to be described in a SM gauge invariant fashion by the $d=6$ operators of Eq.~\eqref{eq:dipole_d6} which, after EWSB, match to the ones of Eq.~\eqref{eq:dipole_UV}. 
These operators give rise to decay rates for the sterile neutrino which, in the limit of vanishing SM lepton masses, read
\begin{align}
& \Gamma_{N\to \nu^i\gamma} = 
\frac{m_N^3}{16\pi}
(2 {\cal C}_{\cal B}^i c_\omega + {\cal C}_{\cal W}^i s_\omega)^2 v^2  \ , \nn \\
& \Gamma_{N\to \nu^i Z} =
\frac{m_N^3}{16\pi}
(2 {\cal C}_{\cal B}^i s_\omega - {\cal C}_{\cal W}^i c_\omega)^2 v^2 
\left(
1-\frac{m_Z^2}{m_N^2}
\right)^2
\left(
1+\frac{m_Z^2}{2m_N^2}
\right)
\ , \nn \\
& \Gamma_{N\to e^{+i} W^-} =
\frac{m_N^3}{16\pi}
{\cal C}_{\cal W}^{i,2}  v^2 
\left(
1-\frac{m_W^2}{m_N^2}
\right)^2
\left(
1+\frac{m_W^2}{2m_N^2}
\right)
\ , 
\end{align}
where the charged current decay rate refers to only of the two charge conjugated channels. 
As regarding the production modes, the situation is more involved with respect to the simplified case of Sec.~\ref{sec:warm} due to $SU(2)_L$ correlations among the various diagrams and the possible presence of all the three $d_\gamma^i$, $d_Z^i$ and  $d_W^i$ operators, see Eq.~\eqref{eq:dipole_UV}. To simplify the discussion and in order to be concrete we will analyse the following two different choices for the UV Wilson coefficients
\begin{itemize}
\item Benchmark I: ${\cal C}_{\cal W}^i=0$ \ ,
\item Benchmark II: ${\cal C}_{\cal W}^i=2 {\cal C}_{\cal B}^i \tan_\omega $ \ .
\end{itemize}

In the first benchmark the sterile neutrino has non-vanishing dipoles only with the neutral EW bosons, while in the second benchmark it retains non zero dipole interactions only with the $\gamma$ and $W$ bosons. We will start by analysing the former case, that shares more similarities with the one studied in Sec.~\ref{sec:warm} in terms of production topologies, moving later to the second benchmark choice.

\subsection{Benchmark I: ${\cal C}_{\cal W}^i=0$}\label{sec:bp_1}

\begin{figure*}[t!]
\begin{center}
       \includegraphics[width=0.482\textwidth]{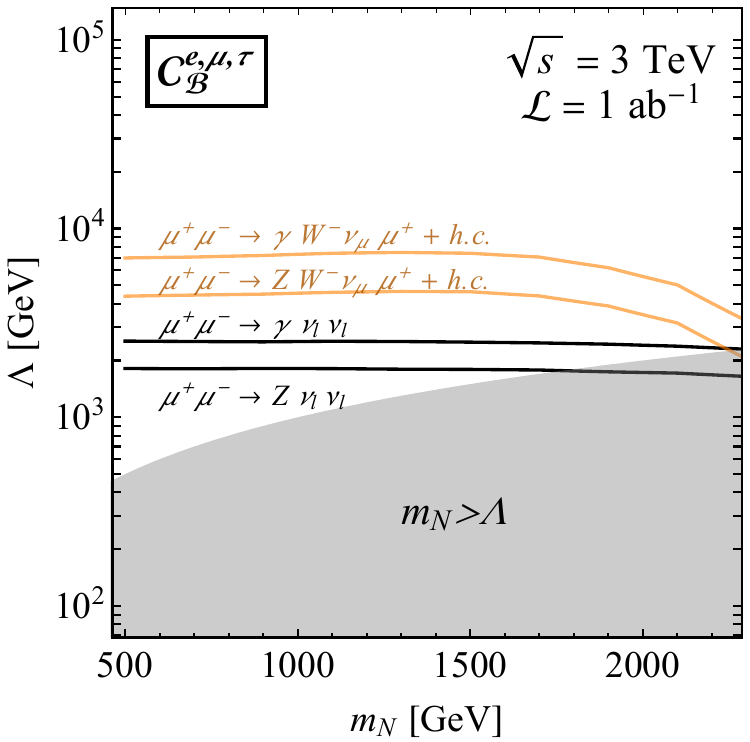}\hfill
        \includegraphics[width=0.482\textwidth]{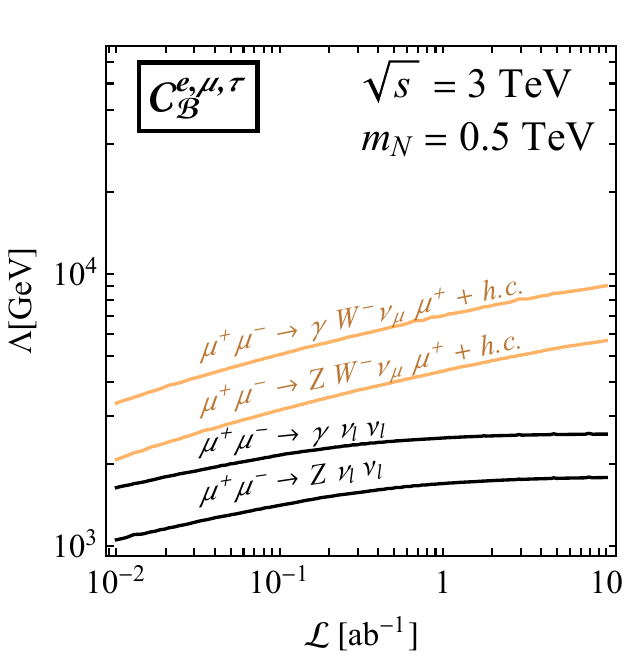}\\
        	\includegraphics[width=0.482\textwidth]{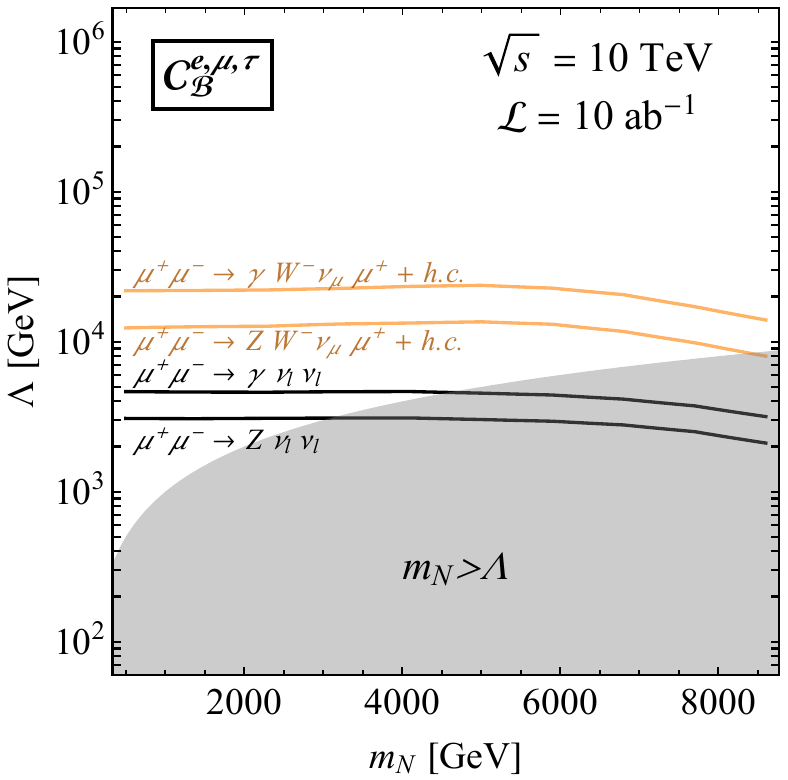}\hfill
 \includegraphics[width=0.482\textwidth]{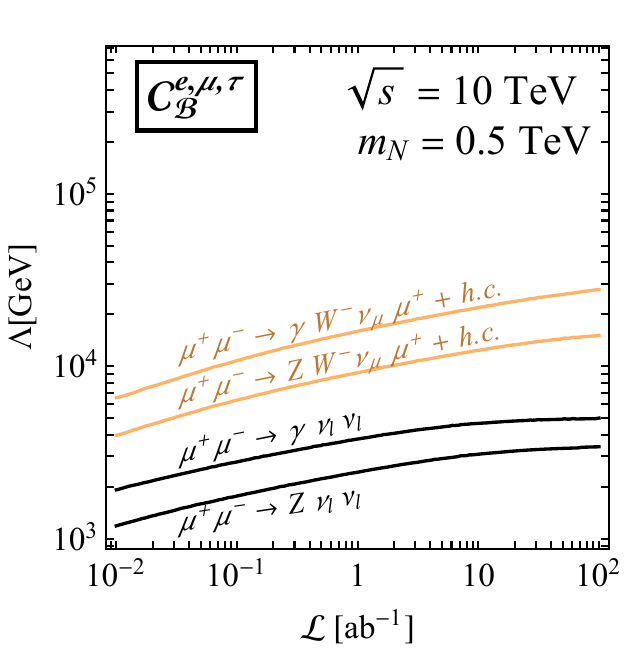}\hfill
 \caption{
Expected sensitivity on the new physics scale $\Lambda$ for the benchmark scenario with ${\cal C}_{\cal W}^i=0$, assuming a single non-vanishing flavor coupling at the time. The black and orange contours show the $s-$channel and $V\nu$ fusion reach. 
In all cases we distinguish between the $N\to \gamma \nu$ and $N\to Z\nu$ decays.
We fix $\sqrt s =3\;$TeV and $\sqrt s =10\;$TeV in the upper and lower panels respectively. In the left plots we show the reach in function of the sterile neutrino mass assuming the target $\mu$C integrated luminosity of Eq.~\eqref{eq:lumi}, while in the right plots we show the results in function of the collected dataset for a fixed sterile neutrino mass. See main text for more details.
}\label{fig:reach_GI_za}
 \end{center}
 \end{figure*}

In this first benchmark, which we label ${\cal C}_{\cal B}^{e,\mu,\tau}$ in Fig.~\ref{fig:reach_GI_za}, the relevant signal topologies are the same of the ones of Sec.~\ref{sec:warm}, with the difference that now both in the $s-$channel and in the  fusion processes the involved EW boson could be either a $\gamma$ or a $Z-$boson. 
The same is true for the sterile neutrino decay, which could produce both $V \nu$ final states with $V=\gamma, Z$. 
We distinguish among the processes where the sterile neutrino decays into $\gamma \nu$ and $Z \nu$, taken to be separate and distinguishable final states.
For the $s-$channel production we consider the background processes  $\mu \mu \to \nu \bar\nu V$, where in the case $V=Z$ we  only focus on its hadronic decay channel and apply the corresponding reconstruction efficiency as discussed in Sec.~\ref{sec:warm}. 
For the  fusion production whose final state is
$\mu \mu \to  \mu W \nu_\mu V$ with $V=\gamma$ similar considerations to the background case of Sec.~\ref{sec:warm} apply, while 
on the other side if $V=Z$, the relevant final state is now $\mu W Z \nu_\mu\,$
and we impose the relevant BRs and reconstruction efficiencies discussed in Sec.\,\ref{sec:warm} for the EW bosons.
Given the similarities with the warm-up analysis, at generator level 
we impose the same cuts
that have been applied in
Sec.~\ref{sec:warm}.

For what concerns the signal cross-section, the relative rates of the $\gamma \nu$ over $Z \nu$ rates are, neglecting phase space effects,  driven by the ratio of the partial widths in these
channels which scales as $ \simeq c_\omega^2/s_\omega^2 \simeq 3$. This reflects into a slightly smaller reach when considering the sterile neutrino decaying into a $Z\nu$ final state for both the $s$-channel and the $V \nu$  fusion topologies, also taking into account that for the $V=Z$ case we only focus on the hadronic final state. Leptonic
decay channels can in principle be combined in order to improve the sensitivity. As regarding the
signal and background kinematic distributions,
 they present similar shapes as compared to the warm-up case of Sec.~\ref{sec:warm}, given the similarity of the topologies. We therefore refrain from showing them and consequently apply the same selection cuts already employed before in the
 warm-up case of Sec.~\ref{sec:warm}. Representative cut-flows for the signal and backgrounds cross-sections are shown in Tab.~\ref{tab:cut_flow_bp1} for the case $\sqrt s= 3\;$TeV.

\begin{table}[h!]
\begin{center}
\scalebox{0.87}{
\begin{tabular}{ c |  c c || c | c c }
\multicolumn{6}{c}{
$\mu$C at $\sqrt s=3\;$TeV~~~${\cal L}=1\;$ab$^{-1}$~~Benchmark I: ${\cal C}_{\cal W}^i=0$~~$N\to\gamma\nu$}\\
\hline
 \multicolumn{3}{c||}{{\bf $s-$channel}} & \multicolumn{3}{c}{{\bf 
Fusion}} \\
\hline
Cut 
& $\sigma_{\rm bkg}\;$[fb]
& $\sigma_{\rm sgn}\;$[fb]
& Cut
& $\sigma_{\rm bkg}\;$[fb]
& $\sigma_{\rm sgn}\;$[fb]  \\ 
\hline
Gen. level 
& $2.0 \times 10^2$
& 32
& Gen. level
&  10
& $ 4.5\times 10^{2}$\\
$p_T^\gamma>900\;$GeV
&   5.9
& 12
& $p_T^\gamma>600\;$GeV,\;$2.44<|\eta^\mu|<7$
&  $1.4 \times 10^{-1}$
&  64
\end{tabular}
}
\vskip 20pt
\scalebox{0.87}{
\begin{tabular}{ c |  c c || c | c c }
\multicolumn{6}{c}{
$\mu$C at $\sqrt s= 3\;$TeV~~~${\cal L}=1\;$ab$^{-1}$~~Benchmark I: ${\cal C}_{\cal W}^i=0$~~$N\to Z\nu$}\\
\hline
 \multicolumn{3}{c||}{{\bf $s-$channel}} & \multicolumn{3}{c}{{\bf 
Fusion}} \\
\hline
Cut 
& $\sigma_{\rm bkg}\;$[fb]
& $\sigma_{\rm sgn}\;$[fb]
& Cut
& $\sigma_{\rm bkg}\;$[fb]
& $\sigma_{\rm sgn}\;$[fb]  \\ 
\hline
Gen. level 
& $2.2 \times 10^{2}$
& $4.3$
& Gen. level
&  8.4
&  64\\
$p_T^Z>900\;$GeV
&   2.6
&   1.6 
& $p_T^Z>600\;$GeV,\;$2.44<|\eta^\mu|<7$
&  $1.2 \times 10^{-1}$
& $9.3$
\end{tabular}
}
\end{center}
\caption{
Cut-flow for the $s-$channel and fusion processes and corresponding backgrounds for the
benchmark scenario with ${\cal C}_{\cal W}^i=0$
at a $\mu$C with $\sqrt s=3\;$TeV. The upper and lower tables refer to the $N\to \gamma \nu$ and $N\to Z \nu$ final state respectively.
For the signal, we fix $m_N=500\;$GeV and $\Lambda\equiv \mathcal{C}_{\cal B}^{-1/2}=1\;$TeV. The yields refer to a single generation of SM leptons, $i=e,\mu,\tau$,  and the  fusion process is only present for the coupling to the $\mu$ flavor.
See main text for the definition of the generator level cuts.
}
\label{tab:cut_flow_bp1}
\end{table}

Our results are then presented in Fig.~\ref{fig:reach_GI_za} for $\sqrt{s}=3\;$TeV, upper plots, and $\sqrt{s}=10\;$TeV, lower plots. In the left panels  we show
the reach on the NP scale $\Lambda\equiv \mathcal{C}_{\cal B}^{i-1/2}$ with $i=e,\mu,\tau$ for both the $s-$channel and $V\nu$ fusion topologies separately, shown as black and orange contours respectively, assuming the 
target integrated luminosities of $1\;$ab$^{-1}$ and $10\;$ab$^{-1}$ for the two COM options.
In the right panels we show instead the reach on $\Lambda\equiv \mathcal{C}_{\cal B}^{i-1/2}$ in function of the collected dataset, for the choice $m_N=500\;$GeV, which is made in order to factor out phase space effects. In all plots we consider a non vanishing projection of the Wilson coefficient on one flavor at the time. As before, the $s-$channel process is sensitive to all three flavor alignments, while the $V \nu$ fusion one only to the case of a dipole coupling with the second generation. Given that the neutrino flavor is not distinguishable, the reach in the $s-$channel case is identical for all three flavors. 

By inspecting the results we can see that for what concerns the $s-$channel production processes the scale $\Lambda$ that can be tested lies at the edge of the validity of the effective description, being close to the chosen COM energy of the $\mu$C. For the case of the $V\nu$ production one has instead to compare the scale $\Lambda$ with the typical momentum exchanged in the process, which is of order of $m_N$. For this reason we also highlight in gray the region where $m_N > \Lambda$.
 It is also important to stress that the normalization chosen for the dipole operators assumes a strongly coupled UV completion. In the cases of weakly coupled UV completions this operator can only arise at loop-level~\cite{Buchmuller:1985jz,Craig:2019wmo}. In this case the operator expansion presents an additional $1/16\pi^2$ suppression and the NP scale that can be tested needs therefore to be rescaled by a $4\pi$ factor, challenging the validity of the effective description in large part of the parameter space also for the fusion process case. All together the best sensitivity is obtained for $i=\mu$, where the $V\nu$ fusion channel allows us to probe scales $\Lambda\sim 10$ TeV with the target integrated luminosities with the reach degrading slowly for smaller collected datasets, which can thus effectively test the $V\nu$ scattering also in the early phases of the $\mu$C operations.
 We also note that existing collider limits do not constraint the considered parameter space, see {\it e.g}~\cite{Magill:2018jla,Fernandez-Martinez:2023phj}.

\subsection{Benchmark II: ${\cal C}_{\cal W}^i=2 {\cal C}_{\cal B}^i \tan_\omega $}

\begin{figure*}[t!]
\begin{center}
	
 \includegraphics[width=0.482\textwidth]{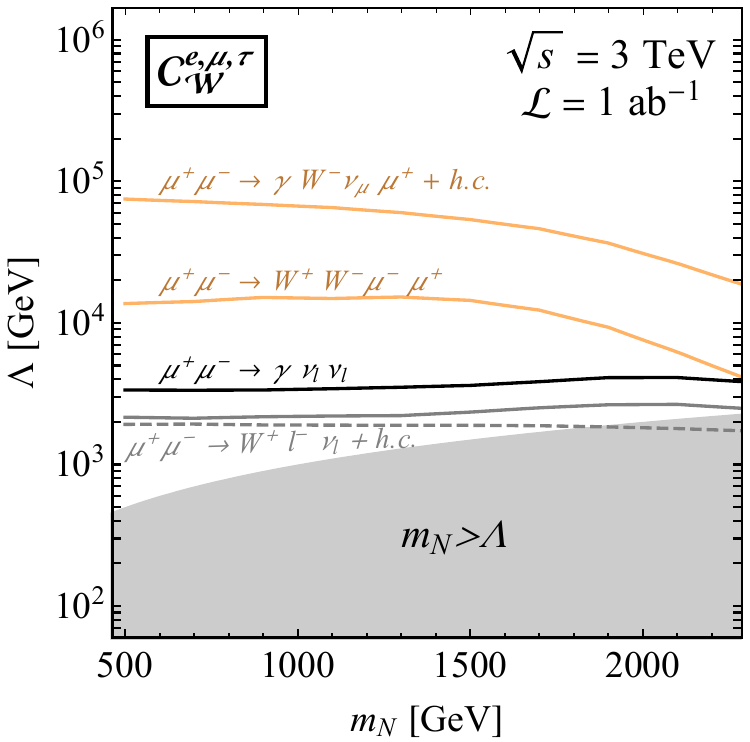}\hfill
 \includegraphics[width=0.482\textwidth]{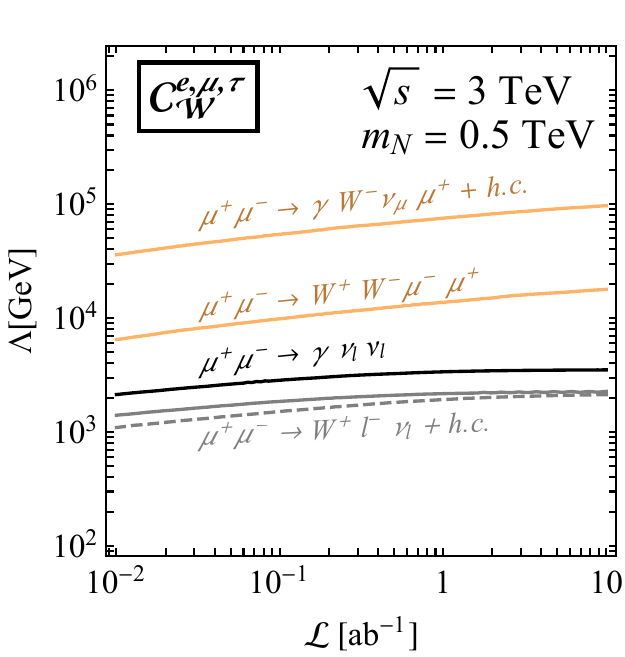}\\
  \includegraphics[width=0.482\textwidth]{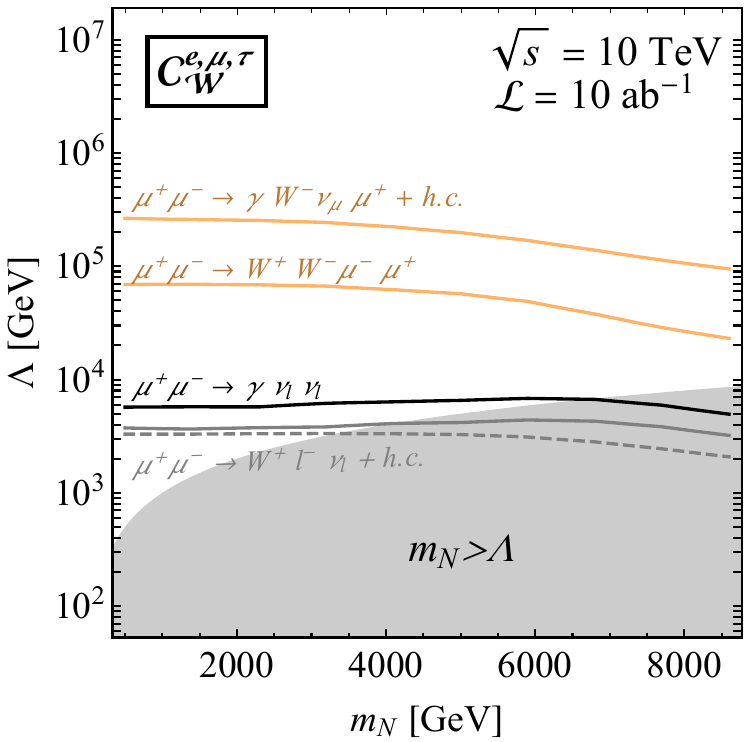}\hfill
 \includegraphics[width=0.482\textwidth]{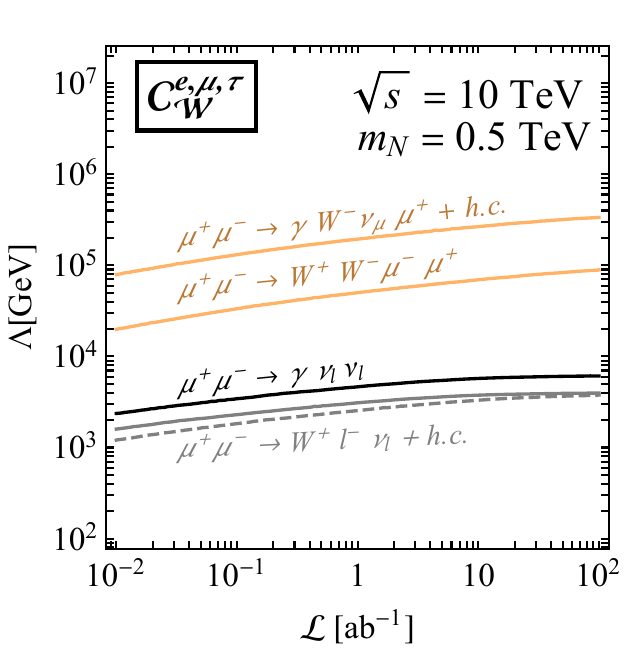}\hfill
 
 \caption{
Expected sensitivity on the new physics scale $\Lambda$ for the benchmark scenario with ${\cal C}_{\cal W}=2 {\cal C}_{\cal B}\tan_\omega$, assuming a single non-vanishing flavor coupling at the time. The black, gray and orange contours show the $s/t-$channel and fusion reach, respectively\color{black}. In all cases we distinguish between the $N\to \gamma \nu$ and $N \to W \ell$ decays. In particular, the gray dashed line represents the $s/t-$channel contributions to the reach for the $N \to W \ell$ final state in the $e$ and $\tau$ flavor cases, while the gray solid lines refers to the reach for the muon flavor. On the other hand, the $s/t-$channel contribution to the $N\to \gamma \nu$ final state is denoted by a black solid line.
We fix $\sqrt s =3\;$TeV and $\sqrt s =10\;$TeV in the upper and lower panels respectively. In the left plots we show the reach in function of the sterile neutrino mass assuming the target $\mu$C integrated luminosity of Eq.~\eqref{eq:lumi}, while in the right plots we show the results in function of the collected dataset for a fixed sterile neutrino mass. See main text for more details. 
}\label{fig:reach_GI_wa_3}
 \end{center}
 \end{figure*}

In this second benchmark, which we label ${\cal C}_{\cal W}^{e,\mu,\tau}$ in Fig.~\ref{fig:reach_GI_wa_3}, the sterile neutrino interacts with the SM leptons also through the $W$ boson, other than with the photon. This allows for different production topologies with respect to the previous case. For what concerns its decay mode the sterile neutrino can now produce a $\gamma \nu$ and $W \ell$ final state with a ratio among the two rates of $s_\omega^{-2} \sim 4$.

The $s-$channel sterile neutrino production proceeds through the same topology of Fig.~\ref{fig:single_prod_IR}. The same final state can however  also be now produced by the $t-$channel exchange of a $W$ boson. Depending on the $N$ decay mode one can have a $\gamma \nu \bar\nu$ or a $W \ell \nu$ final state. 
These are, again, the relevant SM background that we consider in our analysis.

While the $s-$channel topology is present for all flavor assignment of the two Wilson coefficients ${\cal C}^i_{\mathcal{W},\mathcal{B}}$, that we assume to be aligned in flavor space, the $t-$channel is non vanishing only for dipole operators with the second generation of SM leptons. This cross-section, which is effectively a $W\mu$ fusion process, is larger than the corresponding $s-$channel one. However in this case the sterile neutrino tends to be softer and the final
state decay products tend to have kinematic distributions more similar to the ones of the corresponding SM backgrounds, as opposed to the one arising from the $s-$channel topologies. 
For this reason, although at generator level the cross-section for the muon flavor is larger compared to the ones of the electron and tau flavors,
when optimizing the cuts to maximize the reach on the NP scale $\Lambda$ one favors the $s$-channel topology and thus obtain similar reaches for all the three flavors.

At generation level we impose the same cuts as for the case of the previous benchmark for the $\gamma \nu \bar\nu$ final state, while we fix
$|\eta^\ell|<3$ for the $W \ell \nu$ case. 
As mentioned, for final state electrons we choose the same flat 95\% reconstruction efficiency applied to final state muons, again following~\cite{MuonCollider:2022ded}. Final state $\tau$ leptons are instead treated as follows. As for EW gauge bosons we do not simulated their decay, but we instead consider their hadronic decay modes $\tau^\pm \to \pi^\pm \pi^0 \nu_\tau$, 
$\tau^\pm \to \pi^\pm \nu_\tau$ and $\tau^\pm \to \pi^\pm \pi^0 \pi^0 \nu_\tau$ which account for 
approximately 25.5\%, 10.8\% and 9.3\% of the total $\tau$ branching ratio by multiplying for these factors the relevant event yields.
No estimated for the efficiency reconstruction of $\tau$ leptons is yet available~\cite{MuonCollider:2022ded}, and we thus apply a flat 90\% factor following~\cite{Tran:2015nxa}.

Again
at the analysis level we optimise the reach on the NP scale $\Lambda\equiv \mathcal{C}_{\cal W}^{-1/2}$ with $i=e,\mu,\tau$  for the lower sterile neutrino mass that we consider, $m_N=500\;$GeV.
For the $\gamma \nu \bar\nu$ final state we apply the same cuts as for the warm-up case of Sec.~\ref{sec:warm}, while for the $W \ell \nu_\ell$ one we restrict to $|\eta^\ell|<2.44$ and $p_T^W > 900\;$GeV ($p_T^W > 3500\;$GeV) at 
 $\sqrt{s}=3\;$TeV ($\sqrt{s}=10\;$TeV).
Representative cut-flows for both the signal and background cross-sections are reported in Tab.~\ref{tab:cut_flow_bp2} for the case $\sqrt{s}=3\;$TeV, where we only show the results for the muon flavor for compactness. The reaches that we obtain are shown as black  and gray  contours in Fig.~\ref{fig:reach_GI_wa_3} where, as in Fig.~\ref{fig:reach_GI_za}, we present both the sensitivity projections at fixed integrated luminosity varying the sterile neutrino mass as well as varying the integrated luminosity by fixing $m_N=500\;$GeV.  Given the identical production topologies, the reaches for the electron and tau cases are nearly the same, with only minor differences originating from 
the $\tau$ 
reconstruction efficiencies and hadronic BR. 
For this reason, the gray dashed line in Fig.~\ref{fig:reach_GI_wa_3} represents the reach for both the electron and tau flavors, while the gray solid line refers to the muon flavor case.

\begin{table}[h!]
\begin{center}
\scalebox{0.9}{
\begin{tabular}{ c |  c c || c | c c }
\multicolumn{6}{c}{
$\mu$C at $\sqrt s=3\;$TeV~~~${\cal L}=1\;$ab$^{-1}$~~Benchmark II: ${\cal C}^{i}_{\cal W}=2 {\cal C}_{\cal B}^i {\rm tan}\omega$~~$N\to\gamma\nu$
}\\
\hline
 \multicolumn{3}{c||}{{\bf $s-$channel}} & \multicolumn{3}{c}{{\bf 
Fusion}} \\
\hline
Cut 
& $\sigma_{\rm bkg}\;$[fb]
& $\sigma_{\rm sgn}\;$[fb]
& Cut
& $\sigma_{\rm bkg}\;$[fb]
& $\sigma_{\rm sgn}\;$[fb]  \\ 
\hline
Gen. level 
& $2.0 \times 10^2$
& $4.7 \times 10^2$
& Gen. level
& 10
& $2.3 \times 10^6$\\
$p_T^\gamma>900\;$GeV
&    5.9
& 36
& $p_T^\gamma>600\;$GeV,\;$2.44<|\eta^\mu|<7$
&  $1.4 \times 10^{-1}$
&  $8.5 \times 10^5$
\end{tabular}
}
\vskip 20pt
\scalebox{0.9}{
\begin{tabular}{ c |  c c || c | c c }
\multicolumn{6}{c}{
$\mu$C at $\sqrt s=3\;$TeV~~~${\cal L}=1\;$ab$^{-1}$~~Benchmark II: ${\cal C}^i_{\cal W}=2 {\cal C}_{\cal B}^i {\rm tan}\omega$~~$N\to W\ell^i$}\\
\hline
 \multicolumn{3}{c||}{{\bf $s-$channel}} & \multicolumn{3}{c}{{\bf 
Fusion}} \\
\hline
Cut 
& $\sigma_{\rm bkg}\;$[fb]
& $\sigma_{\rm sgn}\;$[fb]
& Cut
& $\sigma_{\rm bkg}\;$[fb]
& $\sigma_{\rm sgn}\;$[fb]  \\ 
\hline
Gen. level 
& $2.8 \times 10^2$
& 95
& Gen. level
&  4.4
&  $5.9 \times 10^4$\\
$|\eta^\ell|<2.44$ 
&  \multirow{2}{1em}{7.4}
& \multirow{2}{1em}{7.5}
& $|\eta^W|<2.44$  $|\eta^\mu|<7$
&  \multirow{2}{1em}{0.12}
&  \multirow{2}{1em}{856}\\
 $p_T^W>900\;$GeV
&   
& 
& $p_T^W>500\;$GeV
&  
&  
\end{tabular}
}
\end{center}
\caption{
Cut-flow for the $s-$channel and Fusion processes and corresponding backgrounds for the
benchmark scenario with ${\cal C}^{i}_{\cal W}=2 {\cal C}_{\cal B}^i {\rm tan}\omega$
at a $\mu$C with $\sqrt s=3\;$TeV. The upper and lower tables refer to the $N\to \gamma \nu$ and $N\to W \ell$ final state respectively.
For the signal, we fix $m_N=500\;$GeV and $\Lambda\equiv \mathcal{C}_{\cal B}^{-1/2}=1\;$TeV. The yields refer to the case of coupling to the muon flavor. See main text for the definition of the generator level cuts.
}
\label{tab:cut_flow_bp2}
\end{table}

Moving then to the fusion production, we observe the following. Firstly, these process again occurs only in the case of a dipole with the second SM lepton generation. Then, the contribution from the $\gamma \nu$ fusion proceeds through the diagram of Fig.~\ref{fig:single_prod_IR} already considered,
producing a $\mu W N$ final state.
On the other side, at lowest order the $W\ell$ fusion process is the $t-$channel contribution already considered together with the $s-$channel contribution. However there are other topologies that can give rise to the $\mu W N$ signal. The dominant, with a rate larger than the $\gamma \nu$ fusion one, 
arises from diagrams where one $\mu$ beam radiates a sterile neutrino emitting a $W$ boson, which fuses into a final $W$ with a photon emitted by the other muon beam. Again we consider the two $N\to \gamma \nu$ and $N\to W \ell$ decays separately.
In the former case the relevant SM background is the one already considered for Benchmak I. 
Also in this case we then keep for the $\mu W \gamma \nu_\mu$ final state the same generation level and optimised analysis level cuts already discussed in Sec.~\ref{sec:warm}.
In the latter case the final state is $WW\mu\mu$ and, to improve the background simulation, we further require at generation level $\Delta R(\mu,\mu)>0.2$.
This cut does not affect the signal rate where the final state muons are well separated. Concerning the event analysis, we restrict both muons pseudorapidities to $|\eta^{\mu}|<7$, with no lower threshold since one muon is produced forward while the other, arising from the $N$ decay, is central. Furthermore, for the $W$ bosons pair we apply $|\eta^{W}|<2.44$ and $p^{W}_T>500\;$GeV ($p^{W}_T>2000\;$GeV) for the $\sqrt{s}=3\;$TeV ($\sqrt{s}=10\;$TeV) analysis. 
Representative cut-flows are reported in Tab.~\ref{tab:cut_flow_bp2}, again for the case $\sqrt{s}=3\;$TeV.
The sensitivity reaches are then shown as orange contours  in Fig.~\ref{fig:reach_GI_za}. We again observe that, while the reaches arising from the $s-$channel processes lie at the edge of the validity of the effective field theory, the reach for the vector fusion signature is  higher than the one obtained in the previous benchmark. With this channel, one can thus probe a large part of the model parameter space already with integrated luminosities smaller than the target one.
Moreover, also when considering a weakly coupled UV completion for the dipole operators, and thus after rescaling the reach on $\Lambda$ by a factor $4\pi$, the fusion process allows to set limits on a larger portion of parameter space complying with the validity of the effective description.
As regarding current bounds  from LHC searches, these have been discussed in~\cite{Magill:2018jla,Fernandez-Martinez:2023phj}. By recasting an 8\,TeV CMS analysis~\cite{CMS:2015loa} a bound of $\Lambda\gtrsim 1\,$TeV for $m_N\simeq 500\,$GeV was set, quickly degrading for heavier $m_N$ masses\footnote{
In extracting the bounds from~\cite{Magill:2018jla} we have considered their benchmark case $a=1$, which is the closest one to our assumption ${\cal C}_{\cal W} = 2 {\cal C}_{\cal B} \tan\theta_\omega$ which matches to $a=1/\sin(2\theta_\omega)$
}. We expect that the corresponding analysis performed at 13\,TeV~\cite{CMS:2018fon} will only be able to increase the bound on $\Lambda$ of at most a few 10\%, therefore not strongly constraining the parameter space considered in this analysis.

\section{Conclusions}
\label{sec:concl}

We have studied the phenomenology of dipole portal operators connecting active and sterile neutrinos at a futuristic $\mu$C, operating at energies between 3\;TeV$-$10\;TeV with integrated luminosities in the multi ab$^{-1}$ range.
These operators allow, for specific flavor alignment of the Wilson coefficients, for two different production modes. 
One proceeding through the exchange of an $s-$channel EW boson and one arising from the fusion of an EW boson with a SM lepton in processes resembling VBF like topologies, which have been proven to be highly relevant for both SM and BSM processes. After having considered as warm-up study 
the photon dipole interaction, in order to highlight the main kinematic features of the signal processes, we have moved to the study of the full-fledged SM gauge invariant embedding of the dipole operators. 
By considering as case study two specific benchmarks choices for the Wilson coefficients of the $U(1)_Y$ and $SU(2)_L$ dipoles, we
have shown that processes initiated by  EW bosons radiated from  incoming muons offer the dominant sensitivity on the effective scale suppressing the dipole interactions, allowing to test new physics in the $\sim 10\;$TeV range when interpreted in terms of strongly coupled UV completions of the dipole operators.

\section*{Acknowledgements}

We thank  Florentin Jaffredo and Paolo Panci for discussions.
DB acknowledges the use of the LXPLUS High Perfomance Computing facility at CERN in the completion of this study.

\appendix

\section{{\tt{MadGraph}} implementation of dipole operators}\label{app:MG}

In order to implement the dipole interactions of Majorana fermions in 
the {\tt UFO}~\cite{Degrande:2011ua} format through the {\tt Feynrules}~\cite{Alloul:2013bka} package
it's important to express the 
interactions of Eq.~\eqref{eq:dipole_d6} through self-conjugate four-component Majorana spinors $\Psi^{\nu,N}_M$ for both the active and sterile neutrinos. Namely 
\be
\Psi^{\nu,N}_M = 
\begin{pmatrix}
\psi^{\nu, N}\\
{\psi^{\nu, N}}^\dag
\end{pmatrix}
\ee
where $\psi$ and $\psi^\dag$ are two-component left-handed Weyl spinors. Then the tensor bilinears for the interactions with the neutral and charged gauge bosons with a generic complex coefficient $a=|a|e^{i \alpha}$
can be written as
\begin{align}
& a \bar \nu_L \sigma^{\mu\nu} N + h.c. = 
|a| \bar \Psi_M^\nu \sigma^{\mu\nu} ( \sin\alpha + i \gamma^5 \cos\alpha) \Psi_M^N \ , \\
& a \bar e_L^i \sigma^{\mu\nu} N + h.c. = 
a \bar e_L^i \sigma^{\mu\nu} P_R \Psi_M^N + h.c. \ ,
\end{align}
where $P_R=(1+\gamma^5)/2$. These are the operators that we have implemented in {\tt Feynrules} with the appropriate choices for the coefficient $a$.

%%%%%%%%%%%%%%%%%
%%%	REFERENCES	   %%%		
%%%%%%%%%%%%%%%%%

\newpage
\bibliographystyle{JHEP}
{\footnotesize
\bibliography{biblio}}

\end{document}